\documentclass{article}
\usepackage{graphicx}
\usepackage{amsmath}

\newcommand{\order}{\mathcal{O}}
\newcommand{\ave}[1]{\left \langle #1 \right \rangle}

\title{A note on a paper by Erik Volz:\\ SIR dynamics in random networks}
\author{Joel C. Miller\\ {\footnotesize \texttt{joel.c.miller.research@gmail.com}}\\ {\footnotesize Harvard School of Public Health}\\ {\footnotesize 677 Huntington Ave}\\ {\footnotesize Boston, MA 02215, USA}}

\begin{document}
\maketitle
\begin{abstract}
  Recent work by Erik Volz~\cite{volz:cts_time} has shown how to
  calculate the growth and eventual decay of an SIR epidemic on a
  static random network, assuming infection and recovery each happen
  at constant rates.  This calculation allows us to account for
  effects due to heterogeneity in degree that are neglected in the
  standard mass-action SIR equations.  In this note we offer an
  alternate derivation which arrives at a simpler --- though
  equivalent --- system of governing equations to that of Volz.  This
  new derivation is more closely connected to the underlying physical
  processes, and the resulting equations are of comparable complexity
  to the mass-action SIR equations.  We further show that earlier
  derivations of the final size of epidemics on networks can be
  reproduced using the same approach, thereby providing a common
  framework for calculating both the dynamics and the final size of an
  epidemic spreading on a random network.
\end{abstract}

\section{Introduction}
Infectious diseases are constrained to spread along the contacts of a
population.  Mathematical models investigating epidemics typically
assume that the contacts occur through mass action
mixing~\cite{kermack,andersonMay}.  However true populations violate
some mass-action assumptions in a manner affecting the epidemic
dynamics.  Recently a number of investigations have been performed
using random
networks~\cite{newman:spread,kenah:second,miller:heterogeneity,babak:finite,meyers:sars}
which allow for a better accounting of mixing in the population.

Unlike mass-action models, random networks allow for the number of
contacts individuals have to remain bounded as the population size
increases.  Thus once an individual infects a contact, the number of
available contacts to infect decreases by a non-negligible amount.
Random networks also allow for more accurate representation of
heterogeneities in the number of contacts compared with mass-action
models.  In a population with heterogeneous contact levels,
individuals with more contacts are preferentially infected early in
the epidemic (and in turn cause more infections), while at the end of
the epidemic the remaining susceptibles tend to have fewer contacts.

A number of analytic results have been found for epidemic probability
or size in random networks, but with only a few exceptions
(notably~\cite{babak:finite,volz:cts_time}), no analytic attention has
been paid to the dynamics of the growth in networks.  However, some
attempts have been made using \emph{pair approximations} which track
the number of joined pairs of individuals with $k_1$ contacts and $k_2$
contacts in each infection state~\cite{eames:pair} (assuming
infection and recovery occur at constant rates).  For a network with
$n$ different degrees, such a model results in $\order(n^2)$ coupled
differential equations.

Recent work by~\cite{volz:cts_time} has shown that it is possible to
investigate the dynamics of epidemic spread on Configuration Model
networks (described below) using a coupled system of only three ODEs
(again assuming infection and recovery occur at constant rates).  The
resulting system has many nonlinear terms, but the number of equations
does not grow with the number of different degrees.  In this note we
derive a single differential equation that can capture the dynamics with
only a single higher order term.  The framework we develop to
calculate the dynamics can also be applied to predicting the final
size of an epidemic in a concise way.  We reproduce earlier results in
this context.

Although our results are equivalent to pre-existing results, we place
previous calculations of epidemic size and epidemic dynamics into a
common framework.  The equations we derive are simpler, and the terms
in the equations are more easily interpreted.  The resulting
calculations for the numbers of susceptible, infected, and recovered
individuals are of comparable complexity to the standard mass-action
$SIR$ equations, but allow for more realistic population interactions.

In section~\ref{sec:framework} we develop the framework for the later
sections.  In section~\ref{sec:dynamics} we apply this framework to
calculating the time course of an epidemic.  In
section~\ref{sec:finalsize} we apply this framework to calculating the
final size of an epidemic.  Finally in section~\ref{sec:discuss} we
discuss the significance of these calculations.

\section{The framework}
\label{sec:framework}

\begin{figure}
\begin{center}
\includegraphics{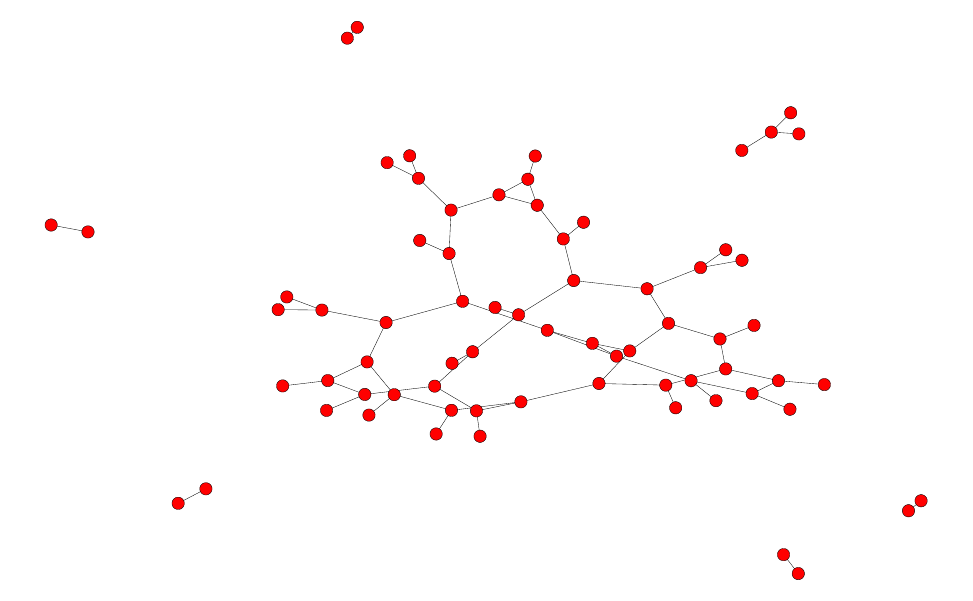}
\end{center}
\caption{A sample Configuration Model network with $70$ nodes.  The degree distribution is chosen such that $P(3)=0.5$ and $P(1)=0.5$.  Thus $\psi(x)=(x^3+x)/2$.}
\label{fig:sample}
\end{figure}

We represent the population by a network.  Each individual is thought
of as a node joined to other nodes by edges through which disease can
spread.  We use Configuration Model (CM)
networks~\cite{newman:structurereview} to model the population.  To
generate a CM network, the \emph{degree} or number of edges of each
node, $k$, is assigned with probability $P(k)$ based on a given degree
distribution.  If the sum of degrees is odd, all degrees are
reassigned until the sum is even.  Then each node is placed into a
list with repetition equal to its degree, the list is randomized, and
each node in position $2n$ ($n=0,1,\ldots$) is connected with the node
in position $2n+1$.  The resulting network constitutes a uniform
choice from the networks with the given degree distribution.  In
general the network may have self-loops or repeated edges.
For degree distributions with finite mean, the impact of this effect
is negligible in sufficiently large networks and we ignore it.  We
define
\[
\psi(x) =
\sum_{k=0}^\infty P(k) x^k \, ,
\]
the probability generating function of the degree distribution.  Note
that $\psi'(1)=\ave{K}$ is the average degree.  An example CM network
is shown in figure~\ref{fig:sample}.  For many important
distributions, $\psi$ takes a simple form; for example, a Poisson
distribution with parameter $\lambda$ has $\psi(x) = e^{\lambda(x-1)}$.

Nodes in the network are assigned to one of three classes:
susceptible, infected, or recovered.  We denote the fraction of the
population in each class by $S$, $I$, and $R$ respectively.  A
susceptible node becomes infected at rate $n\beta$ where $n$ is the
number of infected neighbors it has.  Once infected, a node recovers
at rate $\gamma$.  A recovered node plays no further role in the
spread.  Typically an outbreak is initiated with a single randomly
chosen infected indvidual in an otherwise susceptible population.

We define an \emph{infectious contact} from $v$ to its neighbor $u$ to
be a contact when $v$ is infectious that would cause infection of $u$
if $u$ were susceptible.  Physically this is the transmission of an
infectious dose from $v$ to $u$.  An individual can cause infectious
contact only when infected.  However, an individual can receive an
infectious contact regardless of his/her state, and so an infectious
contact does not necessarily lead to infection.

We use $\theta$ as a measure of the probability that a random edge has not
transmitted an infectious contact.  Its precise definition is subtle,
but important.  To define $\theta$, we choose an edge uniformly from
all edges.  We then choose a direction for that edge, say from $v$ to
$u$.  We refer to $v$ as the \emph{base} and $u$ as the \emph{target}.
We modify the spread of the disease by disallowing infectious contacts
from $u$ to $v$.  Then $\theta(\infty)$ is the probability that there
is never an infectious contact from $v$ to $u$, while $\theta(t)$ is
the probability that at time $t$ there has not been infectious contact
from $v$ to $u$.  If we did not disallow infection from the target
then an infection of $u$ from some other source would in turn make
infection of $v$ more likely, which in turn makes infectious contact
from $v$ to $u$ more likely, and so transmission along different edges
to the same target would not be independent, thereby complicating the
analysis.

Under the assumption that the spread is deterministic, the cumulative
size of an epidemic at a given time is equal to the probability a
randomly chosen node has been infected.  Disallowing infection from
that single randomly chosen node may impact the dynamics \emph{after}
that node is infected, but it does not modify the probability that
that single node has become infected.  Consequently, to calculate the
size at a given time, it suffices to calculate the probability a
randomly chosen node that cannot infect its neighbors has been
infected, or alternately, is still susceptible.

\section{Dynamics}
\label{sec:dynamics}
To calculate the dynamics, we calculate the fraction of the population
that has not yet been infected.  To do this, we look at the
probability that a randomly chosen node is not yet infected at time
$t$.  We choose a random target $u$ and disallow infection from $u$ to
all of its neighbors.  Using $\theta$ as defined above, if the degree
of $u$ is $k$, then the probability that $u$ is still susceptible is
$\theta(t)^k$.  Thus the fraction of susceptibles is
\begin{equation}
S(t) = \sum_{k=0}^\infty P(k) \theta(t)^k = \psi(\theta(t)) \, .
\label{eqn:S}
\end{equation}

To calculate the rate of change of $\theta$, we will need to
know how many of those edges that have not transmitted an infectious
contact have the opportunity to transmit infection at any given time.
That is, we need to know what proportion of all edges have not
had an infectious contact but come from an infected base node.  We set
$\phi$ to be the probability that the base node of an edge is infected
but the edge has not transmitted infection (assuming as for $\theta$
that the target node does not cause infection).  Those edges which
satisfy the definition for $\phi$ are a subset of those which satisfy
the definition for $\theta$.

We derive coupled differential equations for $\theta$ and $\phi$.  
The rate of change in the probability a random edge has not transmitted
infection is equal to the rate at which infection crosses
edges
\begin{equation}
\dot{\theta} = -\beta \phi \, .
\label{eqn:theta}
\end{equation}
An edge no longer satisfies the definition for $\phi$ when infection
crosses the edge or when the base node recovers.  An edge from $v$ to
$u$ begins to satisfy the definition if $v$ becomes infectious.  The
rate at which neighbors become infectious matches the rate at which
neighbors stop being susceptible.  
We use $h(t)$ to denote the probability that a neighbor is
susceptible, so $\dot{\phi} = -(\beta+\gamma) \phi - (d/dt) h(t)$.  

We now find $h(t)$.  A node is more likely to be a neighbor if it has
more contacts~\cite{feld:friends}, and so the probability the neighbor
has degree $k$ is $kP(k)/\ave{K}$.  The neighbor can only be infected
by an edge other than the one from the target node.  Thus
\[
h(t) = \frac{\sum_{k=0}^\infty kP(k) \theta^{k-1}}{\ave{K}} = \frac{\psi'(\theta)}{\psi'(1)} \, .
\]
Thus the neighbor becomes infectious at rate $-(d/dt) h(t)= \beta\phi
\psi''(\theta)/\psi'(1)$.  We finally get
\begin{equation}
\dot{\phi} = \left[-\beta-\gamma + \beta\frac{\psi''(\theta)}{\psi'(1)} \right] \phi \, .
\label{eqn:phi}
\end{equation}
In fact, we can integrate this equation using~\eqref{eqn:theta} to get
\[
\phi = 1 - (1-\theta) - \frac{\gamma}{\beta} (1-\theta) - \frac{\psi'(\theta)}{\psi'(1)} \, .
\]
The term $1-\theta$ represents the probability the edge has
transmitted an infectious contact, the term $(\gamma/\beta)(1-\theta)$
represents the probability that the base node has been infected but
recovered without an infectious contact, and $\psi'(\theta)/\psi'(1)$
represents the probability that the base node is still susceptible.
The complement of all such edges is exactly those edges which have not
transmitted infection but connect to an infected base node.
Consequently we arrive at
\begin{equation}
\dot{\theta} = -\beta \theta + \gamma(1-\theta) + \beta \frac{\psi'(\theta)}{\psi'(1)} \, .
\label{eqn:finaltheta}
\end{equation}

The epidemiological quantity of interest is only rarely the proportion
of edges which have or have not transmitted infection, but rather it
is usually the values of $S$, $I$, and $R$.  We can calculate $S(t) =
\psi(\theta(t))$ directly.  It is not difficult to show that $\dot{R}
= \gamma I$, and conservation of individuals gives $I=1-S-R$.  Consequently, we
can augment~\eqref{eqn:finaltheta} with
\begin{align*}
\dot{R} &= \gamma I \, ,\\
S &= \psi(\theta)\, ,\\
I &= 1-R-S \, .
\end{align*}
to find $S$, $I$, and $R$.

In order to solve our equations, we need to find appropriate initial
conditions.  At the earliest stages, the outbreak grows
stochastically, and so the deterministic equations are not yet
appropriate.  If an epidemic occurs, eventually the outbreak infects a
large number of nodes and then behaves effectively deterministically.
In a sufficiently large population we can assume that deterministic
behavior begins while the proportion infected is still small compared
to the population.

Once the stochastic phase is over, we have $\theta = 1-\epsilon$ with
$\epsilon \ll 1$.  At early time $\epsilon \propto \exp[(-\beta-\gamma
+ \beta \psi''(1)/\psi'(1))t]$ unless $\psi''(1)$ is infinite (which
corresponds to an infinite variance in the degree distribution such as
occurs in some power-law distributions).  For simplicity we assume the
$\psi''(1)$ is finite (if it were not, growth would not be exponential
initially and this calculation would require more attention).  We
define $t=0$ to correspond to a time when the epidemic is sufficiently
large that the outbreak proceeds deterministically, but the proportion
affected is still small.  From the value of $\theta$ we can easily
calculate $S(t) = \psi(\theta(t))$, and thus we can also calculate
$I+R$.

To distinguish the number of current infections ($I$) from recovered
infections ($R$) requires somewhat more effort.  To find the early
behavior for $R$, we note that $I$ and $\epsilon$ are linearly related
at early time, so that $I \propto \exp[(-\beta-\gamma + \beta
\psi''(1)/\psi'(1))t]$.  Then $\dot{R} = \gamma I$ gives $R = \gamma
I/[-\beta - \gamma + \beta \psi''(1)/\psi'(1)]$.  Combined with
$I+R=1-S= 1- \psi(1-\epsilon)$ this gives $R$ at $t=0$.

\begin{figure}
\includegraphics{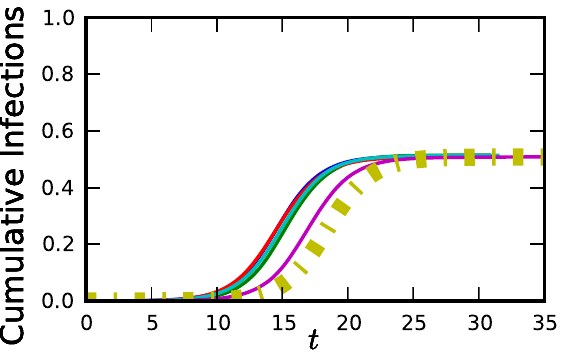}
\hfill\includegraphics{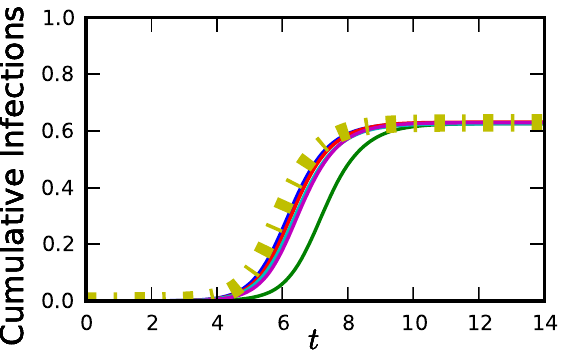}\\
\includegraphics{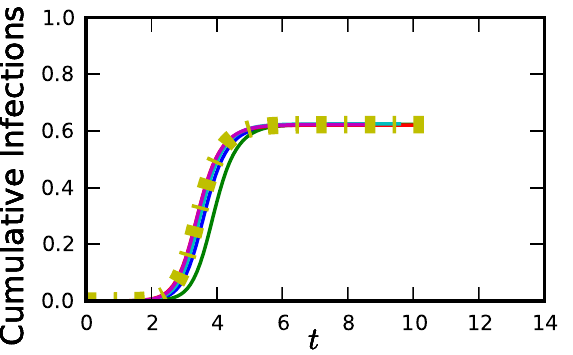}
\hfill\includegraphics{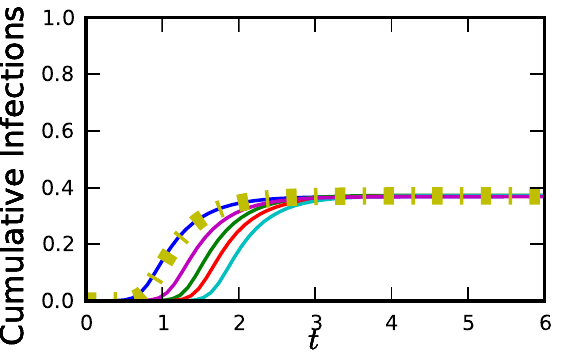}\\
\caption{Plots of cumulative infections $I+R$ against time.  Predicted epidemic
  dynamics (thick, broken curves) and final sizes compared with
  simulations (solid curves) in CM networks  of $500\,000$ individuals
  with $\beta = 1.3$ and $\gamma = 2$.  (a) Every node has degree $4$.  (b) Poisson degree
  distribution of mean $4$. (c) A bimodal
  distribution: $P(1)=5/12$, \ $P(2)=1/12$,
  \ 
  $P(6)=1/12$, and $P(7)=5/12$.  (d) A truncated powerlaw with $P(k)
  \propto e^{-k/40}k^{-2.5}$.}
\label{fig:epidemics}
\end{figure}

We show a comparison of simulation with results calculated using
equation~\eqref{eqn:finaltheta} in figure~\ref{fig:epidemics}.
The results show good agreement, except for time shifts resulting from
stochastic effects in the simulations while the outbreak size is still
small.


\subsection{Discussion}
Equation~\eqref{eqn:finaltheta} contrasts with the original system
of~\cite{volz:cts_time} which uses three equations.  In addition to
the variable $\theta$, the system of~\cite{volz:cts_time} uses $p_I =
\phi/\theta$ (the probability that an edge is connected to an infected
node given that it has not transmitted infection to the target node)
in place of $\phi$ and an additional variable $p_S$ (the probability
that an edge is connected to a susceptible node given that it has not
transmitted infection to the target node):
\begin{align*}
\dot{\theta} &= -\beta p_I \theta\, ,\\
\dot{p}_I &= p_I \left[\beta p_S \theta \frac{\psi''(\theta)}{\psi'(\theta)} - (\beta+\gamma) + \beta p_I\right]\, ,\\
\dot{p}_S &= \beta p_S p_I \left[ 1- \theta \frac{\psi''(\theta)}{\psi'(\theta)}\right] \, .
\end{align*}
We have replaced this system by the single
equation~\eqref{eqn:finaltheta} with only one higher order term.  To see
that these systems are equivalent, we note the $\dot{p}_S$ equation
can be eliminated by observing that the probability the neighbor has
not been infected is $\psi'(\theta)/\psi'(1)$ and so
$p_S=\psi'(\theta)/\theta\psi'(1)$.  Equation~\eqref{eqn:phi} can be
modified by using $\psi'(1)=\psi'(\theta)/\theta p_S$ and $\phi =
\theta p_I$ to arrive at the same $\dot{p}_I$ equation.

\section{Final epidemic size}
\label{sec:finalsize}
We now reproduce some of the earliest results for epidemics on
networks~\cite{miller:heterogeneity,newman:spread,andersson:network}
by calculating the final size of epidemics (under the assumption that
the outbreak does not die out during the stochastic phase).  We can
find this by solving equation~\eqref{eqn:finaltheta} for
$\dot{\theta}=0$.  However, this approach is unnecesarilly specific
and we can easily generalize to disease processes that do not depend
on constant infection and recovery rates by calculating
$\theta(\infty)$ directly rather than through equations for the
intermediate dynamics.  To simplify notation in this section we use
$\theta_\infty$ to represent $\theta(\infty)$ as we are not interested
in the epidemic state at intermediate time.

To calculate the epidemic size, we look for the probability that a
randomly chosen node $u$ is never infected.  If a node has degree $k$,
then the probability that it is never infected is $\theta_\infty^k$.  From
this we get
\begin{equation}
S(\infty) = \sum_{k=0}^\infty P(k) \theta_\infty^k = \psi(\theta_\infty) \, .
\label{eqn:sizeS}
\end{equation}
We must calculate $\theta_\infty$.  We set $T= \int_0^\infty\gamma
e^{-\gamma \tau}( 1-e^{-\beta \tau}) \, d\tau=
\beta/(\gamma+\beta)$.  This is the probability that a randomly
chosen neighbor has an infectious contact with $u$ given that the
neighbor becomes infected.  If $h$ is the probability that the
neighbor does not become infected (given that $u$ does not transmit
infection), the probability of infectious contact is $T(1-h)$.  Thus
the probability of not transmiting is
\[
\theta_\infty = 1-T(1-h)  = 1-T+Th\, .
\]
An argument in the previous section shows that
$h=\psi'(\theta_\infty)/\psi'(1)$, and so $\theta_\infty$ solves the implicit
relation
\begin{equation}
\theta_\infty = 1-T+T\frac{\psi'(\theta_\infty)}{\psi'(1)} \, .
\label{eqn:sizetheta}
\end{equation}
Using~\eqref{eqn:sizeS} and~\eqref{eqn:sizetheta} together gives
$S(\infty)$, and the final size of an epidemic is simply
$1-S(\infty)$.

Note that the ability of a base node to infect a neighbor depends on
duration of infection and whether the base node becomes infected.
Consequently, infectious contacts along different edges out of the
same node are not independent events (they both depend on the base
node's properties).  However, this does not affect our calculations
because infectious contacts along different edges into the same node
are independent events.  If there were variation in susceptibility,
more work would be needed~\cite{miller:heterogeneity}.  Also the
independence assumption will fail if short cycles are not negligible
because infection of one neighbor is correlated with infection of
another.

\section{Discussion}
\label{sec:discuss}
We have shown that calculations for both the final size and the
dynamics of an epidemic on a random network can be placed into a
common framework.  This framework allows us to simplify previous
calculations of the dynamics~\cite{volz:cts_time}.  Our calculations
match closely to simulations, except for time shifts that result from
stochastic effects when the infected population is still small.  Our
model is of similar complexity to the standard mass-action SIR
equations.

The assumption that the network is a Configuration Model network is
central to this derivation.  If there is a tendancy for high degree
individuals to preferentially contact high degree individuals, these
approaches do not directly apply.  Similarly the presence of many
short cycles will also affect these calculations.  When a short cycle
exists, whether or not one neighbor of the target node is still
susceptible may no longer be independent of whether another neighbor
is still susceptible.

\section*{Acknowledgments}
I am grateful to Erik Volz for useful comments.  This work was
developed while preparing a talk for the China-Canada Colloquium on
Modeling Infectious Diseases in Xi'an, China September 2009.  The work
was supported by the RAPIDD program of the Science \& Technology
Directorate, Department of Homeland Security and the Fogarty
International Center, National Institutes of Health.

\end{document}